\newcommand{\Lag}{\mathcal{L}}
\newcommand{\diff}{\textrm{d}}

\newcommand{\rmin}{r_{\text{min}}}

\newcommand{\be}{\begin{equation}}
\newcommand{\ee}{\end{equation}}

\newcommand{\lsim}   {\mathrel{\mathop{\kern 0pt \rlap
  {\raise.2ex\hbox{$<$}}}
  \lower.9ex\hbox{\kern-.190em $\sim$}}}
\newcommand{\gsim}   {\mathrel{\mathop{\kern 0pt \rlap
  {\raise.2ex\hbox{$>$}}}
  \lower.9ex\hbox{\kern-.190em $\sim$}}}

\documentclass[universe,article,submitted,pdftex,moreauthors]{Definitions/mdpi} 
\usepackage{subfig}
\usepackage{graphicx}
\usepackage[export]{adjustbox}
\usepackage[normalem]{ulem}

\firstpage{1} 
\pubvolume{1}
\issuenum{1}
\articlenumber{0}
\pubyear{2024}
\copyrightyear{2024}
 
\datereceived{ } 
\daterevised{ }  
\dateaccepted{ } 
\datepublished{ } 
\hreflink{https://doi.org/}    

\Title{Mimicking wormholes in Born-Infeld electrodynamics}

 \TitleCitation{Mimicking wormholes in Born-Infeld electromagnetism}

 \Author{Jose Beltr{\'a}n Jim{\'e}nez $^{1,\ddagger}$\orcidB{}, Luis J. Garay$^{2,\dagger}$\orcidA{},  and Mar\'ia P{\'e}rez Garrote $^{1,}$* \orcidC{}}
 
 \AuthorNames{Luis J. Garay, Jose Beltr{\'a}n Jim{\'e}nez and Mar{\'i}a P{\'e}rez Garrote}

 \AuthorCitation{J. Beltr{\'a}n Jim{\'e}nez, L.J. Garay,  M.P{\'e}rez Garrote}
 
 \address{$^{1}$ \quad Departamento de Física Fundamental and IUFFyM, Universidad de Salamanca, E-37008 Salamanca, Spain;\\
$^{2}$ \quad Departamento de F\'isica Te\'orica and IPARCOS,
Universidad Complutense de Madrid, 28040 Madrid, Spain;}

\firstnote{jose.beltran@usal.es}
\secondnote{luisj.garay@ucm.es }  
\corres{idu067473@usal.es}

\abstract{ 
We  compute  the  evolution of linear perturbations on top of a background solution of a general nonlinear electromagnetic theory. This evolution can be described in terms of two effective metrics, and we analyse under what conditions they are conformally related, so that they can be regarded as analogue models of non-trivial gravitational fields in the eikonal approximation. This is the  case of Born-Infeld theory. For the  background created by a static point electric charge in the Born-Infeld theory, the effective metric describes a wormhole geometry for light rays. Depending on the impact parameter, incoming light rays are either scattered to infinity or approach the wormhole slowing down their pace until they hit the charge at vanishing speed. The same effective wormhole geometry is obtained for a magnetic monopole and a dyon and we relate it to the duality invariance of Born-Infeld electromagnetism. Finally, we analyse the scalar Dirac-Born-Infeld theory and show that the effective wormhole geometry is not generated by a particle with scalar charge.
}


\begin{document}

\let\svthefootnote\thefootnote
\let\thefootnote\relax\footnotetext{\hspace*{-1.8em} Preprint No.: IPARCOS-UCM-24-056}\let\thefootnote\svthefootnote

\section{Introduction}
Nonlinear electrodynamic models have been extensively studied as extensions of Maxwell electrodynamics with multiple applications (see e.g. \cite{Sorokin:2021tge} and references therein). One of the most notable examples is the Born-Infeld nonlinear extension of Maxwell's theory \cite{Born1933FoundationsOT, foundations} , motivated by the removal of the infinite self-energy of a point charge. A second paradigmatic example was obtained by Heisenberg and Euler \cite{Heisenberg:1936nmg} by integrating out the electron at one loop and showing that strong electromagnetic fields can polarize the vacuum, thereby modifying Maxwell's equations with nonlinear terms. 

The general construction of nonlinear electromagnetic theories are obtained by replacing Maxwell Lagrangian with a nonlinear function of Lorentz invariants of the electromagnetic field. Nonlinear theories that recover Maxwell electromagnetism in the weak-field limit are said to obey the principle of correspondence \cite{osti_4071071}. Some nonlinear theories, such as the Born-Infeld theory, also share some of the symmetries of Maxwell electromagnetism, e.g. invariance under parity transformations and under duality transformations, i.e. Hodge duality rotations \cite{Schrodinger1935,Gibbons:1995cv,Hatsuda:1999ys}. It has been recently shown that conformal and duality invariance of Maxwell's Lagrangian is shared by another family of nonlinear electromagnetisms~\cite{Bandos:2020jsw} that do not satisfy the principle of correspondence. An important feature of nonlinear electromagnetism is that the propagation speed of waves propagating on a non-trivial electromagnetic background depends on both the point and the polarization of the wave \cite{Boillat:1970gw,osti_4071071}. The same physical phenomenon occurs in Maxwell electromagnetism for electromagnetic waves propagating in a medium \cite{LLV8,Born:1999ory}, and this has been used to interpret non-trivial backgrounds in nonlinear electromagnetism as photons propagating in a medium.

Given the non-trivial nature of wave propagation in a nonlinear electromagnetic field, it is possible to develop analogue models of gravity through nonlinear electrodynamics (see e.g. \cite{doi:10.1142/S0217751X98002183,Baldovin:2000xy,PhysRevD.63.103516,novello2002artificial,Goulart:2024ldm}). The fundamental concept behind an analogue model of gravity is to reproduce the propagation of a field in a curved spacetime by means of a field propagating on a background characterized by appropriate space and time-dependent properties. For example, a well-known analogy is phononic propagation on a Bose–Einstein condensate as an analogue for light propagation on a relativistic curved spacetime \cite{Garay:1999sk} (see \cite{Barcelo:2005fc} for a comprehensive review on analogue gravity).

To formulate an analogue model using nonlinear electrodynamics, one introduces linear perturbations around a background electromagnetic configuration. We will compute the quadratic Lagrangian governing the linear dynamics of the perturbations and re-derive the known result that the propagation of the perturbations can be described, in the eikonal approximation, by two effective metrics that depend on the characteristics of the background \cite{Boillat:1970gw,osti_4071071,Novello:1999pg,Obukhov:2002xa,Visser:2002sf}. Moreover, under certain conditions, both metrics can be related by means of a conformal factor. Thus, the propagation of high frequency perturbations in the nonlinear background is analogous to the propagation of light rays according to a single effective metric \cite{Novello:1999pg,Obukhov:2002xa,Visser:2002sf}. Therefore, by appropriately selecting the nonlinear electromagnetic theory and a background configuration, it is possible to construct analogue models for non-trivial gravitational fields. In this work, we will exploit this fact to construct a wormhole geometry within the Born-Infeld theory \cite{foundations,Born1933FoundationsOT}. The possibility of constructing an effective wormhole geometry was recognised in \cite{Baldovin:2000xy} for the nonlinear electromagnetism proposed by Born in \cite{Born:1934ji}, which is very closely related to the Born-Infeld theory \cite{Born1933FoundationsOT, foundations}, but it does not share its exceptional features such as e.g. duality invariance or absence of shock waves. Here, we will perform a thorough analysis of the wormhole geometry within Born-Infeld theory and we will show that, by virtue of its duality invariance, the effective wormhole geometry arises for general dyons, thus explaining why a point-like charge and a magnetic monopole generate the same effective geometry for the linear perturbations.  Furthermore, we will also tackle the case of the its scalar analogue, the Dirac-Born-Infeld theory, and show that the effective wormhole geometry is not generated by a point-like particle with scalar charge.

This work is organized as follows. We will begin in section \ref{sec1} by describing a general nonlinear theory of electromagnetism, and introducing linear perturbations described by a quadratic Lagrangian. This will allow us to obtain two effective metrics that describe the propagation of the perturbations in section \ref{sec2}. Then, we will analyse under which conditions for the background the two effective metrics can be related by a conformal factor, and show that it is not possible to write the conformal factor as the square root of the determinant of a single effective metric. We will focus on Born-Infeld theory in section \ref{Born-Infeld theory}. We will obtain the effective metric for an electric background generated by a point charge in Born-Infeld electrodynamics, which satisfies the characteristics of a wormhole geometry. In section~\ref{Electric wormhole}, we will also study the propagation of null geodesics on the wormhole as a function of the impact parameter. Analogously, we will analyse the dynamics of perturbations in a background generated by a monopole in section \ref{sec ultima} and by a dyon in section \ref{sec dyon}. Lastly in section \ref{Sec DBI}, we briefly consider the scalar Dirac-Born-Infeld theory.
We conclude this work with a summary of the results and conclusions in section \ref{conclusions}.

\textbf{Conventions:} We use the
metric signature $(-,+,+,+)$. The field strength of the four-vector potential $A_\mu(x)$ is $F_{\mu\nu}=\partial_\mu A_\nu-\partial_\nu A_\mu$. The dual of the field strength is given by $\tilde{F}^{\mu\nu}=\frac{1}{2}\epsilon^{\mu\nu\alpha\beta} F_{\alpha\beta}$.
The electric and magnetic fields are defined as $E_i=F_{0i}$ and $B_i=\Tilde{F}_{0i}$, respectively.

\section{Nonlinear electromagnetism}\label{sec1}

We will start by presenting the background equations for a generic nonlinear electromagnetism and the subsequent analysis of the linear perturbations in such background. The resulting equations in the eikonal approximation will be the basis for our analogue of
photons propagating on a wormhole geometry for Born-Infeld electromagnetism.

\subsection{Background dynamics}\label{background}

Let us consider a general theory for a nonlinear electromagnetism in flat spacetime, described by the Lagrangian
\begin{equation}\label{Lag}
    \Lag=K\left(Y,Z\right)+A_\mu J^\mu\:,
\end{equation}
where $Y=-\frac{1}{4}F_{\mu\nu}F^{\mu\nu}$ and $Z=-\frac{1}{4}F_{\mu\nu}\widetilde{F}^{\mu\nu}$ are the two independent Lorentz invariants (the latter only under proper transformations), $K\left(Y,Z\right)$ is an arbitrary function of them, and $J^\mu$ is the current acting as an external source. The two quantities $Y$ and $Z$ can be written in terms of the electric and magnetic fields as $Y=\frac{1}{2}\left(\Vec{E}^2-\Vec{B}^2\right)$ and $Z=\Vec{E}\cdot\Vec{B}$. 
The general Lagrangian \eqref{Lag} reduces to the Maxwell Lagrangian when $K\left(Y,Z\right)=Y$.  

In the nonlinear theory, the Euler-Lagrange equations are
\begin{equation}\label{eom}
    \partial_\nu\left( K_YF^{\mu\nu}+K_Z\Tilde{F}^{\mu\nu}\right)=J^\mu,
\end{equation}
where $K_Y$ and $K_Z$ stand for the corresponding partial derivatives.
In addition to the equations of motion, we also have the Bianchi identity for the field strength tensor \mbox{$\partial_\mu\tilde{F}^{\mu\nu}=0$}, as a consequence of gauge invariance.

\subsection{Perturbations}\label{perturbations}

The objective of this work is to study the propagation of perturbations within a non-trivial electromagnetic background. For this purpose, we consider small perturbations $f_{\mu\nu}(t,\vec{x})$ around a solution of the equations of motion \mbox{$F_{\mu\nu}(t,\Vec{x})$}. The quadratic Lagrangian describing the linear perturbations is
\begin{align} 
    \Lag& =-\frac{1}{4}K_Yf_{\mu\nu}f^{\mu\nu}-\frac{1}{4}K_Zf_{\mu\nu}\Tilde{f}^{\mu\nu}+\frac{1}{4}K_{YZ}f_{\alpha\beta}F^{\alpha\beta}f_{\mu\nu}\Tilde{F}^{\mu\nu}
    \nonumber\\
    &\quad\ +\frac{1}{8}K_{YY}\left(f_{\mu\nu}F^{\mu\nu}\right)^2+\frac{1}{8}K_{ZZ}\left(f_{\mu\nu}\Tilde{F}^{\mu\nu}\right)^2.\label{Lag em f}
\end{align}
In this expression and the following, $K_Y$ and $K_Z$ are the derivatives of $K$ with respect to $Y$ and $Z$, evaluated on the background, and likewise for the higher order derivatives. The quadratic Lagrangian \eqref{Lag em f} clearly shows how the linear perturbations generally propagate in an anisotropic and inhomogeneous background determined by the background electromagnetic field. We can rewrite the quadratic Lagrangian \eqref{Lag em f} in the more apparent form
\begin{equation}
    \Lag=-\frac{1}{8}\Omega^{\alpha\beta\mu\nu}f_{\alpha\beta}f_{\mu\nu}-\frac{1}{4}K_Zf_{\mu\nu}\Tilde{f}^{\mu\nu},
\end{equation}
where we have introduced the background object
\begin{equation}
\Omega^{\alpha\beta\mu\nu}\equiv K_Y\left(\eta^{\alpha\mu}\eta^{\beta\nu}-\eta^{\beta\mu}\eta^{\alpha\nu}\right)-K_{YY}F^{\alpha\beta}F^{\mu\nu}-K_{ZZ}\Tilde{F}^{\alpha\beta}\Tilde{F}^{\mu\nu}-2K_{YZ}F^{\alpha\beta}\Tilde{F}^{\mu\nu}.
\end{equation}
Since we will be mostly interested later in configurations with only an electric or magnetic field in the background, we can set $Z=0$ so that, assuming parity invariance, \mbox{$K_Z = K_{YZ} = 0$}. Among other consequences, the non-trivial background induces a (local) dependence of the propagation speed of the perturbations on both the direction of propagation and the polarization of the wave. This same phenomenon also occurs for light propagation in anisotropic and/or inhomogeneous media in optics \cite{Born:1999ory}, so that we can establish an analogy between nonlinear electrodynamics and standard electromagnetism in a medium whose optical properties are related to $\Omega^{\alpha\beta\mu\nu}$. In this work we will be interested in relating the linear propagation in nonlinear electromagnetism and photons moving on a curved spacetime. The goal will thus be to relate $\Omega^{\alpha\beta\mu\nu}$ with the geometrical properties of some curved spacetime.

Before proceeding to establishing said analogy, let us point out that we will be interested in parity invariant theories, so that they are invariant under the complete Lorentz group. This invariance implies that the Lagrangian $K\left(Y,Z\right)$ must be even in $Z$, and the terms $K_{YZ}$ and $K_Z$ in \eqref{Lag em f} will therefore be odd in $Z$. Thus, for parity invariant theories, odd derivatives with respect to $Z$ will vanish for purely electric or magnetic backgrounds for which $Z = 0$, that will be the backgrounds of interest for us. These derivatives will also vanish for backgrounds having $\Vec{E}\cdot\Vec{B}=0$ but with $\Vec{E}$ and $\Vec{B}$ non-vanishing. We will not consider this possibility in this work, although we will eventually study dyon-like solutions that have $Z\neq0$ in section \ref{sec dyon}.

\section{On the existence of a geometric analogue}\label{sec2} 

In this section, we will analyse the required conditions that permit to reduce the propagation of the linear perturbations described by \eqref{Lag em f} to the propagation of photons on top of some geometry described by a metric. In other words, we will study the requirements for the object $\Omega^{\alpha\beta\mu\nu}$ to admit a factorisation in terms of effective metrics. The effective geometry description in nonlinear electromagnetism has been extensively studied in the literature \cite{Boillat:1970gw,osti_4071071,Novello:1999pg,Obukhov:2002xa,Visser:2002sf}. We will re-derive known results here for completeness. Furthermore, we will seize the occasion to highlight some features that are important for our goal in this work. A caveat of our goal stems from the conformal (or, more precisely, Weyl) invariance of Maxwell electrodynamics in four dimensions. This symmetry suggests that the effective metric, in the cases where it exists, will leave an undetermined conformal factor since the propagation of photons in Maxwell electromagnetism only depends on equivalent classes of conformally related metrics.

\subsection{Effective bimetric theories}\label{Effective bimetric}

The existence of a factorisation of $\Omega^{\alpha\beta\mu\nu}$ is not a property that is guaranteed \textit{a priori}. Furthermore, the existence of a metric description is a stronger requirement because the factorisation has to be a \textit{perfect square} (modulo a conformal factor). We will then commence by attempting to factorise $\Omega^{\alpha\beta\mu\nu}$ in terms of two, in principle different, metrics so the quadratic Lagrangian takes the form \footnote{Although we do not write explicitly the symmetry properties of the factorisation, the obvious (anti-)symmetries of $\Omega^{\alpha\beta\mu\nu}$ will be inherited by the factorisation.}
\begin{equation}\label{ansatz}
    \Lag=-\frac{1}{4}g_1^{\mu\nu}g_2^{\rho\sigma}f_{\mu\rho}f_{\nu\sigma},
\end{equation}
where $g_{1,2}^{\mu\nu}$ are some effective metrics for the perturbations that depend on the background configuration which we will determine in the following.
Note that we could also consider a term of the strict form $f_{\mu\nu}\Tilde{f}^{\mu\nu}$ in the Lagrangian \eqref{ansatz}. However, this term is a total derivative and, hence, would not contribute to the equations for the perturbations so we can safely neglect for our purposes. According to the index structure of $\Omega^{\alpha\beta\mu\nu}$ the metrics can only depend on the background fields in the specific form
\begin{equation}\label{metrics}
    g_i^{\mu\nu}=\mathcal{A}_i\eta^{\mu\nu}+\mathcal{B}_iF^{\mu\rho}F_\rho{}^\nu,
\end{equation}
where $\mathcal{A}_i$ and $\mathcal{B}_i$ are scalars that, in general, depend on the background fields, and hence on the function $K\left(Y,Z\right)$, its derivatives, and the two invariants $Y$ and $Z$. Let us notice that we could have added an additional term $F^{\mu\rho}\tilde{F}_\rho{}^\nu$, but, upon use of the identity $F^{\mu\rho}\tilde{F}_\rho{}^\nu=Z\eta^{\mu\nu}$, we see that it is already included in $\mathcal{A}_i$. Since $Z$ is a pseudo-scalar, the inclusion of $F^{\mu\rho}\tilde{F}_\rho{}^\nu$ would require a pseudoscalar piece inside $\mathcal{A}_i$ which in turn will be related to e.g. $K_{YZ}$.

For the sake of generality, we will not assume $Z = 0$ from the onset and we will only set it to zero at the end. We will then consider the general quadratic Lagrangian \eqref{Lag em f} and compare it to our desired factorisation \eqref{ansatz} and \eqref{metrics}. This comparison leads to five independent equations for the coefficients of the effective metrics that are given by:
\begin{align}
    \mathcal{A}_1\mathcal{A}_2+\left(\mathcal{A}_1\mathcal{B}_2+\mathcal{B}_1\mathcal{A}_2\right)Y-\mathcal{B}_1\mathcal{B}_2Z^2&=K_Y, \label{KY}\\
    \frac{1}{2}\left(\mathcal{A}_1\mathcal{B}_2+\mathcal{B}_1\mathcal{A}_2\right)+2\mathcal{B}_1\mathcal{B}_2Y&=K_{YY}, \label{KYY}\\
    \frac{1}{2}\left(\mathcal{A}_1\mathcal{B}_2+\mathcal{B}_1\mathcal{A}_2\right)&=K_{ZZ}, \label{KZZ}\\
    \mathcal{B}_1\mathcal{B}_2Z&=K_{YZ}, \label{KYZ}\\
    0&=K_Z. \label{KZ}
\end{align}
Equations \eqref{KY}, \eqref{KYY}, and \eqref{KZZ} are obtained by equating the terms proportional to \( f_{\mu\nu}f_{\rho\sigma} \) in \eqref{Lag em f} and \eqref{ansatz}, and therefore, they are also valid when \( Z \) does not vanish in the background. We can solve these first three equations in the general case, obtaining
\begin{align}
    \mathcal{B}_2&=\frac{K_{YY}-K_{ZZ}}{2Y\mathcal{B}_1}, \label{B2}\\
    \mathcal{A}_1&=\frac{\mathcal{B}_1}{K_{YY}-K_{ZZ}}\left(2YK_{ZZ}-\sqrt{\Delta}\right), \label{A1}\\
    \mathcal{A}_2&=\frac{1}{2Y\mathcal{B}_1}\left(2YK_{ZZ}+\sqrt{\Delta}\right) \label{A2},
\end{align}
where we have defined
\begin{equation}\label{delta}
    \Delta=-Z^2\left(K_{ZZ}-K_{YY}\right)^2+2Y\Big[K_Y\left(K_{ZZ}-K_{YY}\right)+2YK_{ZZ}K_{YY}\Big] .
\end{equation}
The fact that we can find solutions for the free functions in \eqref{metrics} for the factorisation shows that the Lagrangian  \eqref{Lag em f} admits the bimetric form \eqref{ansatz} and the two effective metrics will have the expression \eqref{metrics} with the aforementioned coefficients. This is the result that has been reiteratively obtained in the literature by employing different methods \cite{Boillat:1970gw,osti_4071071,Novello:1999pg,Obukhov:2002xa,Visser:2002sf}. The last two equations \eqref{KYZ} and \eqref{KZ} consistently recover that \mbox{$K_{YZ}=K_Z=0$} under the considered assumptions, namely, for $Z=0$. If $Z\neq0$, Eq. \eqref{KZ} should be considered as a consistency equation in the eikonal approximation that will be considered below. This is so because the term \mbox{$K_{Z}f_{\mu\nu}\tilde{f^{\mu\nu}}$} in \eqref{Lag em f} can be written, upon integration by parts and using the identity \mbox{$\partial_\mu\tilde{f^{\mu\nu}}=0$}, as \mbox{$-2\partial_\mu K_Z a_\nu\tilde{f^{\mu\nu}}$}. For high-frequency modes with wave-vector $\vec{k}$, this term will then be suppressed with respect to the kinetic term by a factor $\frac{1}{R_{K}\vert\vec{k}\vert}$ with $R_{K}$ parametrically determined by the scale of variation of the background electromagnetic field. Thus, if the background varies slowly as compared to the frequency of the perturbations so \mbox{$R_{K}\vert\vec{k}\vert\gg1$}, we can safely neglect the term $K_Z$ and, as commented, Eq. \eqref{KZ} is to be interpreted as a consistency condition of our approximation.

\subsection{Geometric theories}\label{simple metric}
After showing how the perturbations can be described in terms of two metrics, a natural question to ask is whether these two metrics can be equal so that the perturbations can be described in terms of a single geometry. We will be interested in studying the propagation of light rays as null geodesics of the would-be single metric. Since these geodesics are conformally invariant, actually we will only need the two metrics in the previous section to differ by a conformal factor. Thus, our next goal will be to determine the conditions that the function of the background $K\left(Y,Z\right)$ must satisfy for the effective metrics $g_{1}^{\mu\nu}$ and $g_{2}^{\mu\nu}$ to be equal, up to a conformal factor.
This is equivalent to the condition that $\mathcal{A}_1/\mathcal{A}_2=\mathcal{B}_1/\mathcal{B}_2$. This means that $\Delta=0$, a condition that has been obtained in e.g. \cite{Novello:1999pg}. For a configuration with $Z=0$, this equation simplifies to
\begin{equation}\label{raiz}
    K_{ZZ}=\frac{K_{YY}K_Y}{K_Y+2YK_{YY}}.
\end{equation}
This is the condition that the background must satisfy for the dynamics of perturbations to be described by a single metric and a conformal factor. In this case, the Lagrangian acquires the form
\begin{equation}\label{newLag}
    \Lag = -\frac{1}{4}\Omega^2 g^{\mu\nu}g^{\rho\sigma}f_{\mu\rho}f_{\nu\sigma},
\end{equation}
where
\begin{equation}\label{eff}
    g^{\mu\nu}= \eta^{\mu\nu}+K_{ZZ}\Omega^{-2}F^{\mu\rho}F_\rho{}^\nu,
    \qquad
    \Omega^2=\frac{2YK_{ZZ}^2}{K_{YY}-K_{ZZ}}.
\end{equation}
We can obtain more restrictions by imposing further requirements such as duality invariance. In the absence of charges and currents, Maxwell's equations are invariant
under SO(2) duality rotations, whose conserved charge implies the conservation of photon helicity~\cite{duality}. Duality invariance is also a property shared by a family of nonlinear electromagnetic theories, including Born-Infeld theory \cite{Aschieri:2008zz}. At first order in perturbation theory, duality invariance implies a condition on the derivatives of the Lagrangian evaluated on the background as shown in \cite{Jimenez:2022mih}:
\begin{equation}\label{duality}
    K_{ZZ}=\frac{1}{2Y}\left(K_Y-\frac{1}{K_Y}\right).
\end{equation}
By imposing this invariance in \eqref{raiz}, we obtain an ordinary differential equation in the variable $Y$
\begin{equation}\label{ec}
     2YK_{YY}=K_Y^3-K_Y,
\end{equation}
which implies that $\Omega^2=1/K_Y$.
This equation must be satisfied for any background with $Z=0$ that preserves both parity invariance and duality invariance.
We can integrate \eqref{ec} to obtain
\begin{equation}\label{solu}
    K(Y,0)=\frac{2}{A}\sqrt{1-YA}+C,
\end{equation}
with $A$ and $C$ being constants of integration. 
If $Y$ is small we can expand in Taylor the square root in \eqref{solu} and obtain
\begin{equation}
    K(Y)\simeq -Y+\frac{2}{A}+C.
\end{equation}
Thus, Maxwell electrodynamics is recovered upon Taylor expansion for \( Y A \ll 1 \), as one would expect.
 
We have analysed under which conditions for the background the dynamics of perturbations can be described by a metric and a conformal factor. We may now inquire whether we can express the conformal factor as the square root of the determinant of a single effective metric. This would be the necessary condition for the geometric description to work perfectly for the entire dynamics of the perturbations and not only for the ray propagation.

In view of the form of the Lagrangian, a  natural approach is to impose that  it has the form
\begin{equation}\label{conform}
    \Omega^2g^{\mu\nu}g^{\alpha\beta}=\sqrt{|h|}h^{\mu\nu}h^{\alpha\beta}.
\end{equation}
However, we will see that this cannot be achieved. Let us write
\begin{equation}\label{metric2}
    g^{\mu\nu}=|h|^{n/2}h^{\mu\nu},\qquad
    \Omega^2g^{\mu\nu}=|h|^{(1-n)/2}h^{\mu\nu}.
\end{equation}
where \( n \) is a real number. Taking determinants in \eqref{metric2}, we obtain $\Omega^4 |\det(g^{\mu\nu})|=1$. 
This is a nontrivial condition for the background. If we consider the particular case of    a purely electric or purely magnetic background  and  calculate the determinant of the metric \eqref{eff},
 this condition becomes
\begin{equation}
    K_{YY}-K_{ZZ}=2YK_{YY}K_{ZZ}.
\end{equation}
Recall that condition \eqref{raiz} for the background must hold so that the two effective metrics are related by a conformal factor. These two conditions together imply that \( K_Y = 1 \) and hence that $\Omega^2=1$, i.e. the trivial solution to our problem are unimodular theories. A trivial solution is of course Maxwell's electromagnetism, which falls outside the class of non-linear electromagnetism theories.

This result shows that the theory effectively describing the dynamics of perturbations in the nonlinear background will not be, in general, invariant under conformal transformations. In fact, this invariance is lost when the constraint \eqref{conform} is imposed on the determinants of the effective metrics. Consequently, the propagation of perturbations in a nonlinear background effectively describes the dynamics of photons in curved spacetime only in the regime where light rays are described as null geodesics. We will elaborate on this result in the next section.

\subsection{Geometrical optics approximation}

Nonlinear electromagnetic fields effectively provide a background that behaves as a curved spacetime for the propagation of perturbations. However, perturbations do not evolve in an effective metric strictly speaking, they do so up to a conformal factor.
In the limit of geometrical optics, where wavelengths are much smaller than the characteristic scale of variations of the background curvature (characteristic length of the gradients of the background field in this analogue model), light rays propagate along null geodesics \cite{Stephani:2004ud}, which are conformally invariant. Therefore,  the conformal factor is not relevant to our analysis but only the conformal structure. Let us analyse this limit in our case.

The source-free equations of motion for the perturbations derived from Lagrangian~\eqref{Lag em f}, particularized to $Z=0$, can be written as follows:
\begin{equation}\label{motion}
    \partial_\nu\left(\Omega^{\alpha\beta\mu\nu}\partial_\alpha a_\beta\right)=0,
\end{equation} 
where $a_\mu$ is the perturbation of the vector potential that inherits the gauge symmetry \mbox{$a_\mu\rightarrow a_\mu+\partial_\mu f$} for an arbitrary function $f$.

In the geometrical optics approximation (i.e. for short wavelengths), we can express the vector potential $a_\mu$ of the perturbations by introducing a rapidly varying phase $\theta$ and a slowly changing amplitude $\alpha_\mu$ \cite{Misner:1973prb}
\begin{equation}\label{geo}
    a_\mu=\text{Re}\left(\alpha_\mu e^{-i\theta}\right).
\end{equation}
The equations of motion enable us to derive the dispersion relation for the perturbations. Indeed, if we define the wave vector as $k_\mu=\partial_\mu\theta$ and introduce the vector potential \eqref{geo} into the equation of motion \eqref{motion}, we obtain at lowest order in the wavelength the equation
\begin{equation}\label{first order}
    \alpha_\beta k_\alpha k_\nu \Omega^{\alpha\beta\mu\nu}=0.
\end{equation}
Following the analysis of reference \cite{Visser:2002sf}, let us think of the previous equation as an eigenvalue problem $\alpha_\nu M^{\nu\mu}=0$ for the matrix $M^{\nu\mu}=k_\alpha k_\beta\Omega^{\alpha\nu\mu\beta}$. A necessary and sufficient condition for the eigenvalue problem to have non-trivial solutions is $\det\left(M^{\nu\mu}\right)=0$. It is convenient to adopt the temporal gauge $\alpha_0=0$, so that the eigenvalue problem splits into two equations $\alpha_iM^{i0}=0$ and $\alpha_iM^{ij}=0$. Thus, 
the equation \eqref{first order} admits a non-trivial solution only if
\begin{equation}\label{dispersion}
    \det\left(M^{ij}\right)=0.
\end{equation}
Equation \eqref{dispersion} provides the dispersion relation, i.e. the frequency $k^0$ as a function of the wave vector $k^i$, for light propagating in our non-trivial background. 

By explicitly writing the form of the matrix \( M_{ij} \), it can be easily seen that the determinant has the following structure
\begin{equation}
    \det\left(M^{ij}\right)=(k_0)^2\mathcal{P}_4(k),
\end{equation}
where $\mathcal{P}_4(k)$ is a homogeneous fourth-order polynomial in the wavevector $k_\mu$. Thus, in the most general case, there will be four dispersion relations corresponding to the four roots of the polynomial $\mathcal{P}_4(k)$. However, as we have seen in section \ref{Effective bimetric}, the aforementioned polynomial can be factored into two quadratic forms, i.e., two effective metrics \cite{osti_4071071,Novello:1999pg,Obukhov:2002xa,Visser:2002sf}
\begin{equation}
    \mathcal{P}_4(k)=\left(g_1^{\mu\nu}k_\mu k_\nu\right) \left(g_2^{\alpha\beta}k_\alpha k_\beta\right).
\end{equation}
Under more restrictive conditions, given by equation \eqref{raiz}, these two metrics are equal up to a conformal factor, yielding a single quadratic dispersion relation $g^{\mu\nu}k_\mu k_\nu=0$. This analysis is completely equivalent to the study we have conducted in the previous sections~\ref{Effective bimetric} and~\ref{simple metric}.

\section{Born-Infeld theory}\label{Born-Infeld theory}

Up to this point, we have been working with a general nonlinear electromagnetic theory. In this section, we will specialize the nonlinear electromagnetic theories outlined in \eqref{Lag} to the Born-Infeld theory \cite{foundations,Born1933FoundationsOT}. In Maxwell theory of electromagnetism, the electric field created by a point charge diverges at the position of the particle. This results in an infinite Lagrangian and self-energy of the point particle.  In order to eliminate this divergence, Born and Infeld introduced a new nonlinear theory of electromagnetism, modifying the action function of Maxwell theory into \cite{foundations,Born1933FoundationsOT}:
\begin{equation}\label{lag BI}
    K(Y,Z)=\lambda^4\left(\sqrt{1-\frac{2}{\lambda^4}Y-\frac{Z^2}{\lambda^8}}-1\right),
\end{equation}
where \( \lambda \) is a constant with units of energy. From \eqref{lag BI} we can recover the usual Maxwell action when the electromagnetic field is small compared to the energy scale $\lambda$. Born-Infeld theory stands out among general nonlinear electromagnetism due its exceptional properties (see e.g. \cite{Boillat:1970gw,BOILLAT19729,Tseytlin:1999dj,Gibbons:2001gy,Gibbons:2000xe,Russo:2022qvz,Jimenez:2022mih}). 

The remarkable property of the Lagrangian \eqref{lag BI} for our purposes here is that it automatically satisfies the condition \eqref{raiz} for the background in which the two effective metrics are related by a conformal factor, as can be checked by direct substitution. The exceptional nature of Born-Infeld theory featuring one single effective metric has been repetitively highlighted in the literature (see e.g. \cite{Boillat:1970gw,Novello:1999pg,Obukhov:2002xa,Gibbons:2000xe}). It is also well-known that the Born-Infeld theory is duality invariant~\cite{Schrodinger1935,Gibbons:1995cv, Hatsuda:1999ys}, so it is not surprising that the Lagrangian also satisfies equation \eqref{duality} for $Z=0$, and hence equation \eqref{ec}: the Born-Infeld Lagrangian \eqref{lag BI} corresponds to the solution \eqref{solu} with $A=2/\lambda^4$. Since the background \eqref{lag BI} satisfies equation \eqref{raiz}, we can assert that it is possible to construct two effective metrics that are related through a conformal factor. The Lagrangian for the perturbations in this case can be written as \eqref{newLag} with $\Omega^2=1/K_Y$, and the effective metric as
\begin{equation}\label{ggg}
g^{\mu\nu}=\eta^{\mu\nu}+\frac{\lambda^{-4}}{1-2Y/\lambda^4}F^{\mu\rho}F_\rho{}^\nu.
\end{equation}
In the regime where the eikonal approximation is valid, light rays will effectively propagate through an electromagnetic nonlinear background according to the previous metric. The question that we want to answer next is what type of geometry is described by this metric, i.e., what type of geometry is experienced by photons in Born-Infeld electromagnetism. 

\section{Electric wormhole}\label{Electric wormhole}

In this section, we will specialize to the case of perturbations propagating on the electric field generated by a static charged point-like particle in Born-Infeld electrodynamics. In this case, the field equations can be readily solved due to spherical symmetry and staticity, and the resulting electric field is \cite{foundations}:
\begin{equation} \label{campoelec}
    \Vec{E}=\frac{q}{4\pi r_0^2}\frac{1}{\sqrt{1+(r/r_0)^4}}\hat{r},
\end{equation}
where $q$ is the charge of the particle and $r_0:=\lambda^{-1}\sqrt{\frac{|q|}{4\pi}}$ is the radius that parametrically controls when the nonlinearities become relevant. Here we observe explicitly that the electric field remains finite below the radius $r_0$ and it saturates to the value $\lambda^2$ at the position of the particle. Note that the electric field acquires its usual profile \mbox{$\Vec{E}\simeq\frac{q}{4\pi r_0^2}\hat{r}$} when \mbox{\( r \gg r_0 \)}, and is suppressed with respect to the Maxwellian behaviour for \mbox{\( r \ll r_0 \)}. For this reason, the scale $r_0$ is sometimes referred to as the screening radius, as we will do in the following.

As we are interested in studying the propagation of null geodesics, we will use the metric with a convenient conformal factor. Substituting the expression \eqref{campoelec} for the electric field into \eqref{ggg}, we obtain the following representative of the conformal class of effective metrics
\begin{equation}\label{elemento2}
    \diff s^2=-\diff t^2+\diff r^2+h^2(r)\diff\Omega^2,
\end{equation}
where
\begin{equation}\label{h function}
    h(r)=r\left[1+\left(\frac{r_0}{r}\right)^4\right]^{1/2}.
\end{equation}
This function goes as \mbox{$h(r)\simeq r$} when \mbox{\( r \gg r_0 \)} (well outside the screened region) so the effective metric is the usual Minkowski metric in spherical coordinates, as it should since this regime recovers Maxwell’s solution. However, it presents a minimum at the screening radius and it grows again (as \mbox{$h\simeq r_0^2/r$}) inside the screened region. This behaviour suggests that the metric \eqref{elemento2} describes a wormhole geometry. 

This same geometry was obtained in \cite{Baldovin:2000xy} for a nonlinear electromagnetism described by
\be
 K(Y)=\lambda^4\left(\sqrt{1-\frac{2Y}{\lambda^4}}-1\right),
 \label{eq:Blag}
\ee
that closely resembles the Born-Infeld theory \eqref{lag BI}. In fact, they are equivalent for configurations with $Z=0$. The above Lagrangian was in fact used by Born \cite{Born:1934ji} in his first attempt to regularise the self-energy of the electron before constructing \eqref{lag BI} together with Infeld \cite{foundations,Born1933FoundationsOT}. Thus, the theory \eqref{eq:Blag} could be referred to as {\it precursor} Born-Infeld theory or, even better, simply Born theory. Despite the close relation to Born-Infeld theory, the Lagrangian \eqref{eq:Blag} does not share the unique properties of \eqref{lag BI}. In particular, it is not duality invariant, and this has important consequences as we will see below. Let us notice that, although our purely electric background has $Z=0$ and, hence, both theories have the same purely electric background solution, the perturbations are, in principle, sensitive to the $Z$-dependence of the Lagrangian (the quadratic Lagrangian does depend on derivatives with respect to $Z$) so the effective geometry could have been affected. We find however that this is not the case and both theories generate the effective metric \eqref{ggg} for the electric field produced by a point-like charge.

The metric~\eqref{elemento2} is said to describe a (traversable, Lorentzian) wormhole if the following two conditions are satisfied~\cite{bronnikov2021general}:

\begin{itemize}
    \item First, the area of the sphere of coordinate radius \( r \) must have at least one minimum. We will call \( r_\text{min} \) the value of $r$ where this is achieved. Then, the area of the spheres must be greater than this minimum on both sides of this $r_\text{min}$. For the metric \eqref{elemento2}, the area of the sphere of coordinate radius $r$ is
\begin{equation}\label{area}
    A(r)=4\pi h^2(r)=4\pi r^2\left[1+\left(\frac{r_0}{r}\right)^4\right].
\end{equation}
By minimizing this function, we obtain that the area reaches a minimum value at \mbox{\( r_{\text{min}} = r_0 \)}, as can be seen in Figure \ref{fun area}. Thus, \mbox{\( h(r_0)=\sqrt{2}r_0 \)} corresponds to the area radius of the throat of the wormhole, and \( r_0 \) to its proper radial coordinate.
\begin{figure}
    \centering
    \includegraphics[width=8cm, height=5.5cm]{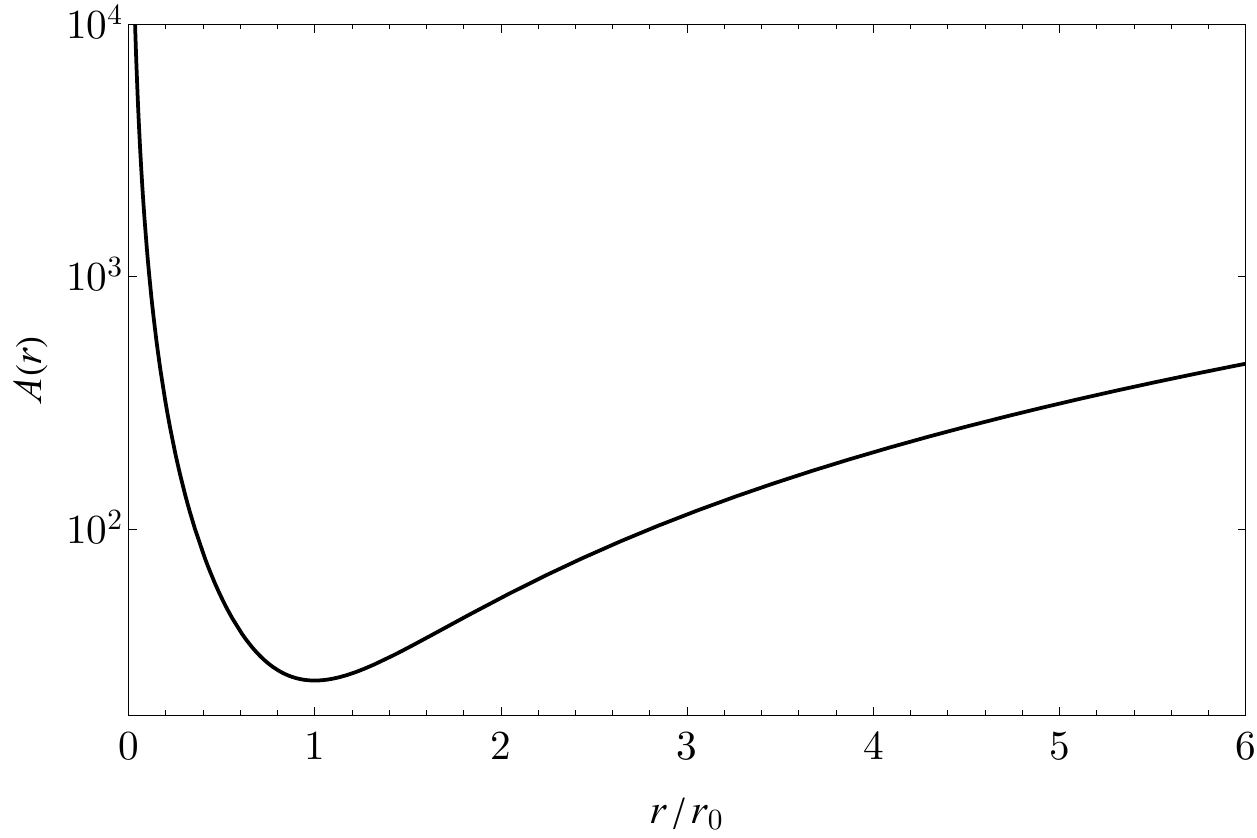}
    \caption{Area of the spheres given by equation \eqref{area} as a function of the coordinate radius \(r/r_0\). The behaviour of this area function shows how the geometry can be interpreted as a wormhole.}
    \label{fun area}
\end{figure}
    \item Second, the functions accompanying $\diff t^2$ and $\diff r^2$ in \eqref{elemento2} must be regular and positive over a range of \( r \) containing the throat and values of \( r \) on both sides of the throat such that \( h(r) \gg h(r_\text{min}) \). Since our metric \eqref{elemento2} is already written in terms of the proper radial coordinate, in which the function accompanying $\diff r^2$ is unity, this condition is trivially satisfied.
        In addition, in these coordinates, the condition for the wormhole to be asymptotically flat is
\begin{equation}
    \left|\diff h/\diff r\right|\rightarrow1\:\:\:\text{as}\:\:\: r\rightarrow\infty,
\end{equation}
which expression \eqref{h function} satisfies, as can be directly checked.
\end{itemize}

Therefore, the metric \eqref{elemento2} describes the geometry of an asymptotically flat wormhole. As we have mentioned, when $r\gg r_0$, we recover the usual Minkowski metric in spherical coordinates. On the other hand, the area function diverges in the limit $r\to0$. In fact, the curvature and Kretschmann scalars of the effective metric are given by
\begin{align}
    R&=-\frac{2\left[5+9\left(r/r_0\right)^2\right]}{r^2\left[1+\left(r/r_0\right)^4\right]^2},\\
    R_{\mu\nu\rho\sigma}R^{\mu\nu\rho\sigma}&=\frac{12\left[3+14\left(r/r_0\right)^4+27\left(r/r_0\right)^8\right]}{r^4\left[1+\left(r/r_0\right)^4\right]^4},
\end{align}
so they as diverge as $1/r^2$ and $1/r^4$, respectively, in the limit \( r \rightarrow 0 \). Consequently, \( r = 0 \) is a curvature singularity. Note that the divergence is smoother than in the case of the Schwarzschild metric \cite{Carroll:2004st}. This divergence can be understood as a result of the pointlike structure of the charge that creates the electric field. It would be possible to regularize this divergence by considering a spherical charge with a certain non-zero radius. However, the Born-Infeld structure near the origin makes the solution more regular than other situations~\cite{alam}.

In general, electromagnetic waves propagating on a medium will have different propagation speeds, which depend on the polarization of the wave in the background. However, in Born-Infeld electrodynamics it is possible to describe the dynamics of perturbations using a single metric, which is equivalent to obtaining a single propagation speed for electromagnetic waves that does not depend on the polarisation, i.e., there is no birrefringence~\cite{osti_4071071,Boillat:1970gw,Gibbons:2000xe,Visser:2002sf,Russo:2022qvz}. The propagation speed will however depend on the direction of propagation and we can read it from the effective metric \eqref{elemento2}. Thus, waves travelling along a radial direction propagate at the Maxwellian speed of light, while waves propagating along angular directions propagate at the speed \mbox{$c_\Omega^2=r^2/h(r)^2$}, in agreement with previous results~\cite{Jimenez:2022mih}. Let us notice that we obtain one single propagation speed for angular directions, which is again a consequence of having one single metric. The angular propagation speed approaches the Maxwellian speed of light in the asymptotic region \mbox{$r\to\infty$} (as it should because we recover the Maxwellian regime in that region), while it becomes subluminal upon entering the screening region \mbox{\( r< r_0 \)}. Well inside the screening radius, we have \mbox{$c_\Omega^2\simeq \left(r/r_0\right)^4$} so that photons propagate arbitrarily slow as they approach the origin. This means that it takes them an increasingly larger time to travel a given proper distance. A consequence of this slower propagation of photons is that angular diameter distances become larger, which can be interpreted as spheres becoming increasingly bigger beyond the throat of the wormhole (see Figure \ref{fun area}). We could be rightly concerned with the fact that the propagation speed becomes zero in the limit \mbox{\( r \rightarrow 0 \)} and this could induce a strong coupling problem near the position of the particle. This is another way of interpreting the related issue of having diverging curvature scalars for the effective metric. Of course, one can argue that this feature by itself does not represent a problem from a physical point of view because the eikonal approximation will cease being valid before reaching the origin. The regime of validity of the eikonal approximation can be roughly estimated by comparing the scale over which the propagation speed varies, namely $r$, and the wavelength  of the perturbation, i.e.,  it will be valid as long as the wavelength is much smaller than $r$. In the asymptotic region this poses no restriction but it obviously implies a breakdown of the approximation as we get closer to $r=0$. Beyond the eikonal approximation, we need to take into account the wave nature of the fluctuations. However, the propagation speed of the fluctuations beyond the eikonal approximation is still given by $c_\Omega^2$ so that does not resolve the potential strong coupling problem. Again, one could argue that effects from the internal structure of the source (that is modelled as point-like) will be relevant so that our approach will not be valid arbitrarily near the origin. An interesting possibility that might allow to push the regime of validity to smaller scales without hitting the strong coupling regime is to invoke higher order derivative operators so the dispersion relation is no longer quadratic and the leading order term comes from e.g. $k^4$ corrections as it occurs in the ghost condensate~\cite{Arkani-Hamed:2003pdi}.
 
\subsection{Null geodesics}\label{Study of geodesics}

Now that we have established the analogue model of a wormhole as the geometry seen by linear perturbations (in the eikonal approximation) propagating on the electric background generated by a point-like charge, let us study in some detail such a propagation. In this case, the trajectories of light rays are described by the null geodesics of the effective metric \eqref{elemento2}. As usual, spherical symmetry implies conservation of the angular momentum, which restricts the motion of trajectories to a plane. Therefore, we will consider, without loss of generality, the equatorial plane \(\theta = \pi/2\).
Let us note that the metric \eqref{elemento2} evaluated in the equatorial plane has two conserved quantities, namely, it possesses two Killing vectors $\partial_\phi$ and $\partial_t$, $\phi$ being the azimuthal angle. Therefore, the quantities \mbox{\( E\equiv\Dot{t} \)} and \mbox{\( \ell\equiv h(r)^2\Dot{\phi} \)} are conserved, and we will refer to them as energy and angular momentum, respectively. We denote with a dot the derivative with respect to the affine parameter $\tau$.
 
On the other hand, by applying the null vector normalization condition \mbox{$g_{\alpha\beta}\Dot{x}^\alpha\Dot{x}^\beta=0$}, which can be associated with the conserved quantity corresponding to the invariance under translations of the parameter $\tau$ in the action, we obtain the following equation for the null geodesics:
\begin{equation}\label{tiempo}
    E^2=\Dot{r}^2+V(r),\qquad V(r):=\frac{\ell^2}{r^2+r_0^4/r^2},
\end{equation}
where we have used the expressions for energy and angular momentum. The effective potential resembles a centrifugal potential, except for a correction proportional to \( r_0^4 \) which dominates at small distances. We can find the maximum value of the potential by solving \mbox{$\diff V(r)/\diff r=0$}, obtaining that, at the radial coordinate \mbox{$r=r_0$}, it reaches the maximum value \mbox{$V(r_0)=\ell^2/2r_0^2$}. This effective potential was also derived in \cite{Baldovin:2000xy}.

\begin{figure}
    \centering
    \includegraphics[width=8cm, height=5.5cm]{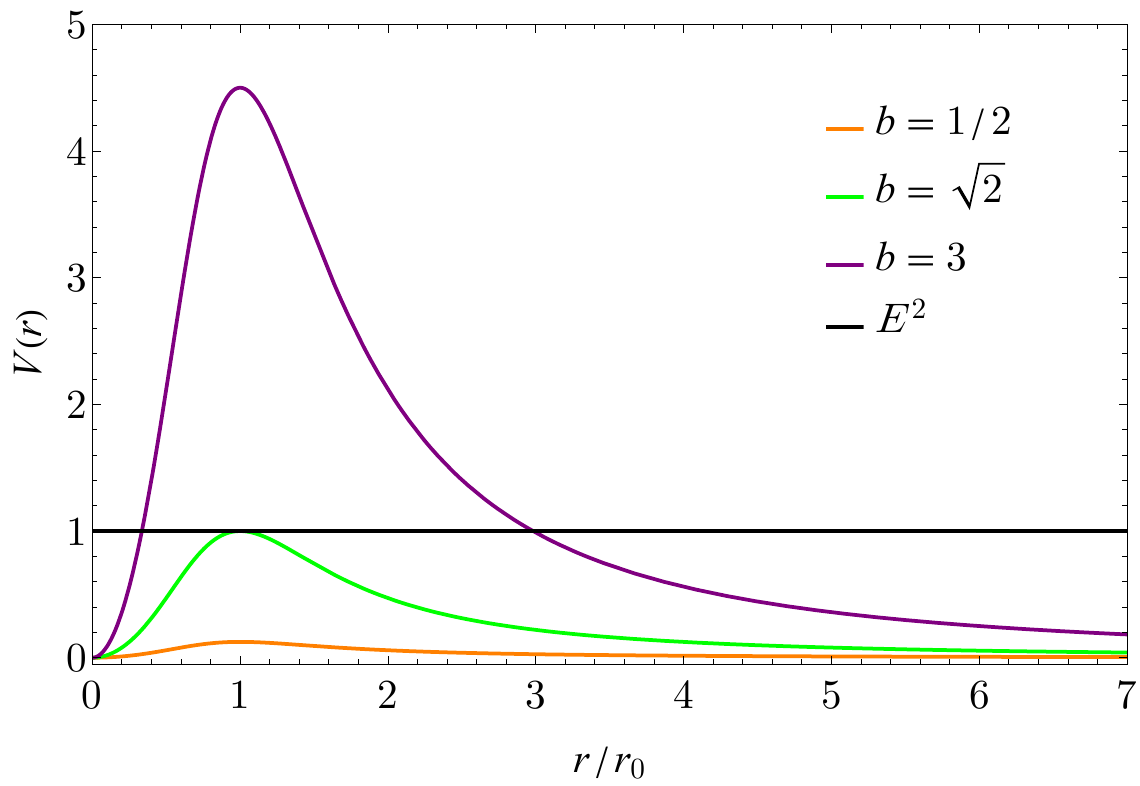}
    \caption{The effective potential \eqref{tiempo} for different values of impact parameter \( b \).}
    \label{pot}
\end{figure}

The behaviour of the orbits depends on the reduced impact parameter \mbox{\( b\equiv |\ell/\left(Er_0\right)| \)}. In Figure \ref{pot}, we plot the effective potential for various values of \( b \). Let us consider a light ray that initiates its trajectory from \mbox{\( r\gg r_0 \)}, and descends towards the wormhole throat. The turning point is located where \mbox{$V(r)=E^2$}, since it is where the radial velocity becomes zero. If \mbox{\( b < \sqrt{2} \)}, there is no turning point, and the ray falls down the throat of the wormhole. On the contrary, if \mbox{\( b > \sqrt{2} \)}, there exists a turning point as we can observe in Figure \ref{pot}. In this case, the ray reaches a certain radius of maximum approximation $r_{\text{min}}$ and then scatters off towards infinity. Finally, the case with the critical impact parameter \mbox{\( b_{\text{crit}} = \sqrt{2} \)} corresponds to rays that enter and remain in circular orbit  at the radial coordinate \( r_0 \), that is, at the throat of the wormhole. However, this orbit is unstable since a small perturbation in the impact parameter leads to the ray either dispersing to infinity or collapsing to~\( r = 0 \). 

In view of this result, we can analyse whether the rays falling through the wormhole reach the region $r=0$ in a finite time. For this purpose, we can integrate the radial velocity in expression \eqref{tiempo} from the throat to the centre, resulting
\begin{equation}
    \tau =\frac{r_0}{E}\int^{1}_0\frac{\diff\rho}{\sqrt{1-\frac{b^2}{\rho^2+1/\rho^2}}},
\end{equation}
where 
$\rho=r/r_0$. We explicitly observe that the arrival time increases with the impact parameter and is finite if \mbox{\( b < \sqrt{2} \)}. Indeed, by numerical integration, we obtain that the descent time is \mbox{\(\tau = 1.03r_0/E\)} when \mbox{\(b = 1/2\)}, and \mbox{\(\tau = r_0/E\)} when \mbox{\(b = 0\)}. Thus, trajectories with nonzero impact parameter require more time to reach \mbox{$r=0$} than radial free fall trajectories, as one would expect. 

\begin{figure}
    \centering
    \captionsetup[subfloat]{labelformat=empty}
    \subfloat[]{
    \includegraphics[width=7cm, height=5.5cm]{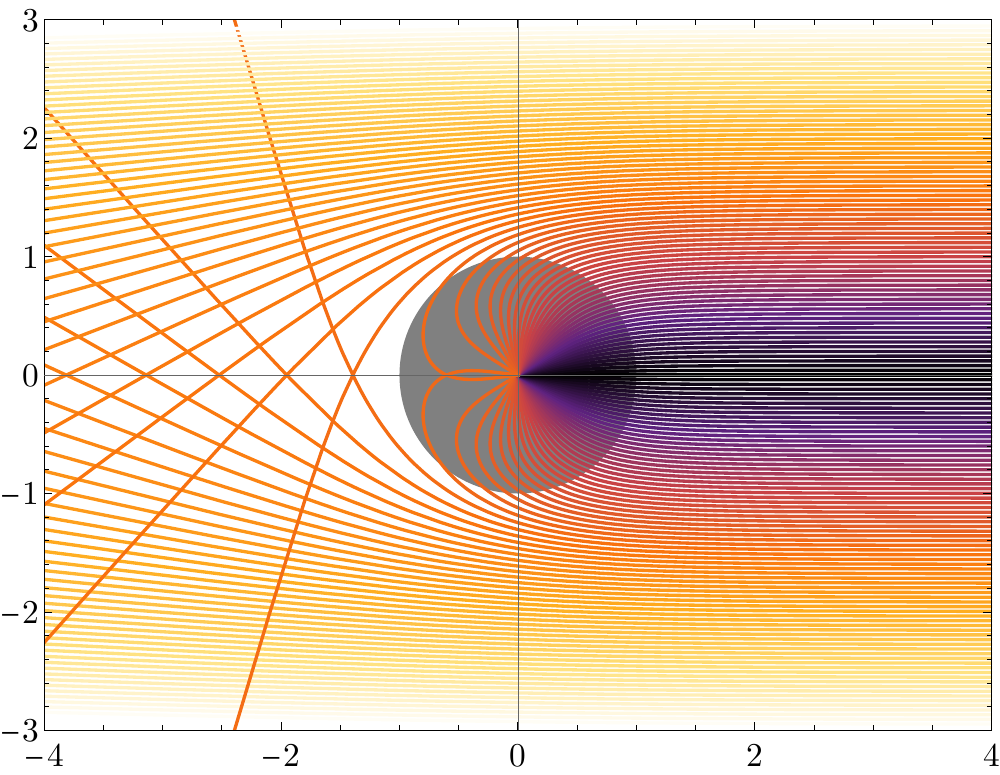}
    \label{tra1}}
    \subfloat[]{
    \includegraphics[width=5.5cm, height=5.5cm]{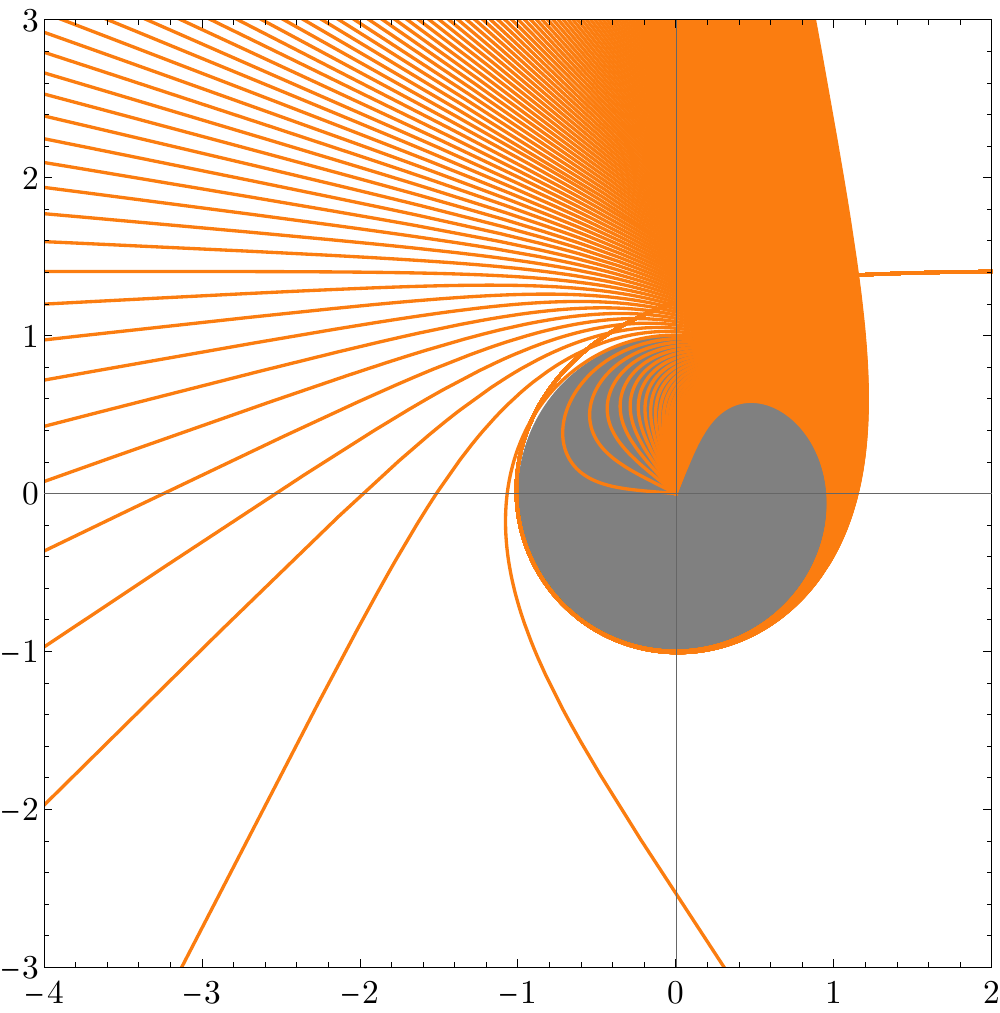}
    \label{WH zoom}}
    \caption{{\bf Left:} Trajectories of an incident homogeneous beam of rays from the right. The gray circle corresponds to the screened region whose edge determines the throat of the wormhole. {\bf Right:} Detail of the trajectories for incident rays with impact parameter very close to the critical value $b=\sqrt{2}$ that, as explained in the main text, determines whether the rays fall towards the centre. We have chosen impact parameters within the range $b=1.41579\pm  10^{-5}$ and everything is in units of $r_0$.}
    \label{trayectorias}
\end{figure}

Differentiating equation \eqref{tiempo} yields
the null geodesic equation for the radial coordinate in the equatorial plane
\begin{equation}
    \Ddot{r}=\frac{\ell^2\left(1-r_0^4/r^4\right)}{r^3\left(1+r_0^4/r^4\right)^2},
\end{equation}
which is more amenable to numerical integration. Together with the conservation of angular momentum \mbox{$\Dot{\phi}=\ell/(r^2+{r_0^4}/{r^2})$}, we can integrate the system to obtain the trajectories of null geodesics. This is what we have done in Figure \ref{trayectorias}, where we show a congruence of null geodesics with different impact parameters. The screened region is represented with a gray circle and its edge corresponds to the throat of the wormhole. The congruence describes geodesics that start at \mbox{\(r \gg r_0\)} and propagate towards the particle. We can clearly identify three regions depending on the value of the impact parameter. When \mbox{\(b \gg \sqrt{2}\)}, the light rays surpass the wormhole throat without substantially deviating from their initial trajectory. As we approach the throat, yet with \mbox{\(b > \sqrt{2}\)}, the lensing effect on the trajectories becomes increasingly significant. Trajectories with an impact parameter \mbox{\(b < \sqrt{2}\)} fall down the throat towards the centre of the wormhole in a finite time. As we have analysed previously, geodesics with \mbox{\(b = \sqrt{2}\)} approach the unstable circular orbit at \mbox{$r=r_0$} from infinity. Any small perturbation will cause the rays to either fall into the centre of the wormhole or escape with a scattering angle as we can see in the right panel in Figure \ref{trayectorias}. We can see how rays passing very close to the critical value but above it suffer from a huge deflection, and they can even go around the throat multiple times before escaping to infinity. Furthermore, the outgoing beam of rays undergoes a large dispersion. On the other hand, rays with impact parameter close to the critical value but smaller can also go around the throat multiple times before falling to the centre. In both cases however the rays either escape to infinity or fall to the centre very rapidly without spiralling.

\begin{figure}
    \centering
    \captionsetup[subfloat]{labelformat=empty}
    \subfloat[]{
    \includegraphics[width=3cm, height=5.5cm]{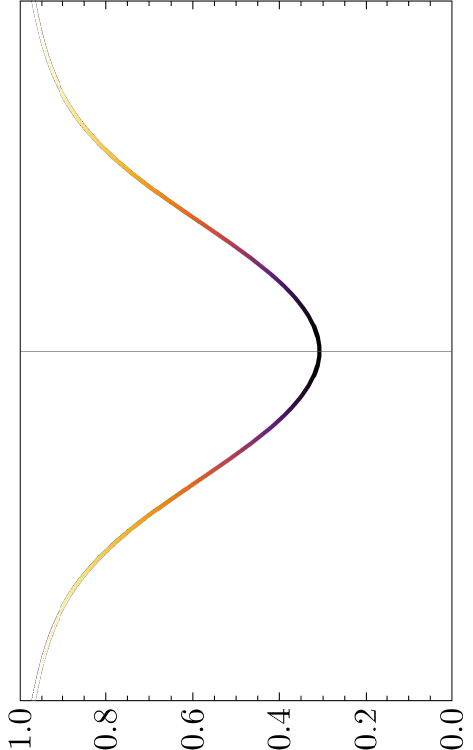}}
    \subfloat[]{
    \includegraphics[width=5.5cm, height=5.5cm]{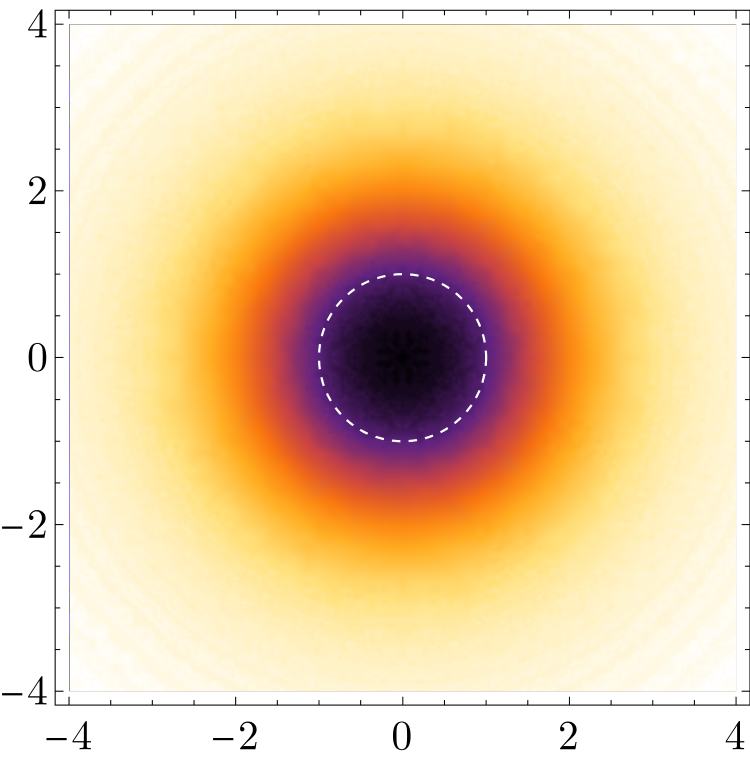}}
    \subfloat[]{
    \includegraphics[scale=0.5]{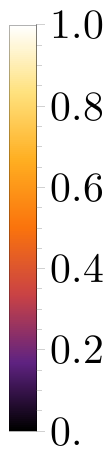}}
    \caption{{\bf Right:} In this Figure we show the image projected on a screen placed at a distance $4r_0$ (we have chosen units of $r_0$ on the screen) from the origin when the particle is illuminated with a wide homogeneous beam. In practical terms, we solve the geodesic equations for a congruence of geodesics starting at a distance $30r_0$ from the particle, and we define the intensity on the screen as the density of trajectories that we compute by binning in bins of size 0.1. We represent the wormhole throat as a white dashed circle. {\bf Left:} Intensity profile received on the screen. The profile is normalized with respect to the maximum intensity value.}
    \label{fig:Image}
\end{figure}

A natural question that we could ask is what is the image that an incident beam of rays would form on a screen placed at some distance from the particle in the outgoing region. The answer to this question is illustrated in Figure \ref{fig:Image} for a screen placed at a distance $4r_0$ behind the throat of the wormhole. We consider an incident homogeneous beam with a transversal cut as the one shown in the left panel of Figure \ref{trayectorias}. From this Figure we can understand and interpret the image shown in \ref{fig:Image}. If there were no particle, we would obtain a featureless image corresponding to the constant intensity image of the ingoing beam. However, the presence of the particle generating the wormhole geometry produces features in the image. For the rays passing far from the screened region, the effect is small and we obtain the constant intensity region far from the centre of the screen. Since the rays with impact parameter smaller that $\sqrt{2}$ fall towards the centre, those rays do not reach the screen and we might expect to see a shadow at the centre of the screen. However, rays with impact parameter close to the critical value undergo significant lensing and we observe a penumbra with a non-vanishing intensity, as we can see in the left panel in Fig.~\ref{fig:Image}. In our treatment, we are are assuming that the rays that fall towards the centre do not propagate any further. This is obviously incorrect and, as explained above, the ray (or eikonal) approximation upon which our analysis is based ceases being valid before reaching the origin. A proper treatment of those rays would require going to higher orders, which is beyond the scope of this work. Finally, let us notice that there are some  rays (those with impact parameter very close to the critical value) that will also illuminate regions of the screen arbitrarily far from the projection of the screened region. However, the intensity of the image formed by those rays is negligible with respect to the incident intensity and, hence, cannot be observed.

The numerical integration of the geodesic equations has permitted us to thoroughly analyse the trajectories of rays. We can however do better than numeric and obtain an analytical expression for an important quantity like the deflection (or lensing) angle $\Delta\phi$, defined as the angle between the incident and outgoing trajectories. We can compute the deflection angle by integrating the trajectory equation

\be
\frac{\diff \phi}{\diff r}=\frac{\dot{\phi}}{\dot{r}}=\frac{1}{h(r)\sqrt{\frac{h^2(r)}{h^2(\rmin)}-1}},
\ee
to obtain
\be\label{exact}
\Delta\phi=2\int_{\rmin}^\infty\diff r\frac{1}{h(r)\sqrt{\frac{h^2(r)}{h^2(\rmin)}-1}}-\pi,
\ee
where we have used the symmetry of the trajectory to integrate only until the maximum approach of the ray, hence the factor 2, and we have subtracted $\pi$ that corresponds to the non-deflected case. As usual, we introduce the variable \mbox{$u=r_\text{min}/r$} and we   use the equation determining the maximum approximation radius
\be
b^2=h^2(r_{\text{min}})/r_0^2.
\label{eq:lrmin}
\ee
Using these variables, the deflection angle can be computed as
\be
\Delta\phi=2\int_0^1\frac{\diff u}{\sqrt{1-u^2}}\left[\frac{1+\alpha}{1+u^2\left(u^2-1\right)\alpha-\alpha^2u^6}\right]^{1/2}-\pi,
\ee
where the dimensionless parameter 
\be
\alpha=\left(\frac{r_0}{r_\text{min}}\right)^4
\ee
controls the correction to the undeflected trajectory of the trivial background. 
For small angles, we can Taylor expand the integrand up to first order in $\alpha$ to obtain
\be
\Delta\phi\simeq2\int_0^1\frac{\diff u}{\sqrt{1-u^2}}\left[1+\frac12\big(1+u^2-u^4\big)\alpha
+\mathcal{O}(\alpha^2)\right]-\pi.
\ee
These integrals can be analytically computed and we obtain
\be
\Delta\phi\simeq\frac{9\pi}{16}
\alpha
+\mathcal{O}(\alpha^2).
\ee
The fact that $\alpha$ scales with the fourth power of \mbox{$r_0/r_\text{min}$} makes the series in $\alpha$ converge quickly so this first order approximation is valid for relatively small values of $\rmin$, i.e., trajectories passing close to the screening radius and, thus, having relatively large deflection angles. In fact, the first order correction 
is accurate to better than 1\% for $\rmin\gtrsim 2r_0$. In any case, it is not difficult to calculate higher order corrections:
\be
\Delta\phi=\frac{9\pi}{16}
\alpha 
+\frac{73\pi}{1024}
\alpha^2 
+\frac{2609\pi}{16284}
\alpha^3 
+\mathcal{O}
(\alpha^4) 
.
\ee

\begin{figure}
    \centering
    \includegraphics[width=8cm, height=5.5cm]{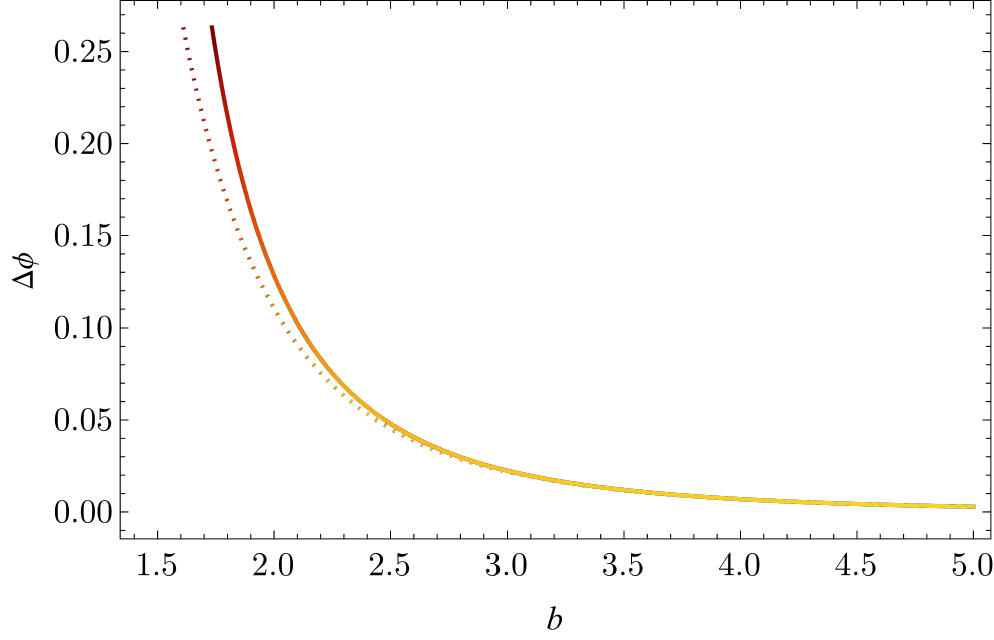}
    \caption{Deflection angle $\Delta\phi$ as a function of the impact parameter $b$. The analytical expression to first order \eqref{LO} has been represented by a dashed line, while the solid line represents the numerically integrated expression \eqref{exact}.}
    \label{deflex}
\end{figure}

Eq. \eqref{eq:lrmin} relates $\alpha$ and the impact  parameter $b$, i.e. \mbox{$b^2=\alpha^{-1/2}(1+\alpha)$.} We then see that the difference between $b^2$ and $\alpha^{-1/2}$ will only contribute to   higher orders. This allows us to express the deflection angle in terms of the impact parameter to lowest order: 
\be\label{LO}
\Delta\phi\simeq\frac{9\pi}{16}\frac{1}{b^4}+\mathcal{O}(b^{-8}).
\ee
We plot in Figure \ref{deflex} the deflection angle $\Delta\phi$ as a function of the impact parameter $b$. We observe that the first-order expression \eqref{LO} (dashed line in Figure \ref{deflex}) aligns well with the exact solution \eqref{exact} (solid line in Figure \ref{deflex}) even for trajectories with impact parameters relatively close to the critical value \mbox{$b_\text{crit}=\sqrt{2}$}. We see that the deflection angle has a vertical asymptote at the critical value \( b = \sqrt{2} \) that corresponds to the rays that stay orbiting in the throat of the wormhole. Near the throat, trajectories scatter significantly, even possibly completing a loop, as can be seen in Figure \ref{trayectorias}. Therefore, the deflection angle is greater the closer we get to \mbox{$b=\sqrt{2}$}. Outside the throat (to the right of the vertical asymptote in Figure~\ref{deflex}), the deflection angle decreases and asymptotically tends to zero, since when the impact parameter is much larger than the throat radius, null geodesics cross the wormhole without deviation.

\section{Magnetic monopole}\label{sec ultima}

Having studied the propagation of perturbations in an electric background, we now consider, for completeness, a magnetic field generated by a static monopole located at the origin, identical to that in Maxwell theory~\cite{PhysRevD.61.085014}
\begin{equation}\label{monopole}
    \Vec{B}=\frac{\mu}{4\pi r^2}\hat{r}.
\end{equation}
Note that this magnetic field trivially satisfies the Born-Infeld equations of motion. By applying the same procedure as in the purely electric case, we derive the following effective metric for null geodesics propagating in a magnetic background generated by a monopole~\eqref{monopole} 
\begin{equation}
    \diff s^2= -\diff t^2+\diff r^2+\tilde h^2(r)\diff \Omega^2,
\end{equation}
where
\begin{equation}
    \tilde h(r)=r\left[1+\left(\frac{\tilde r_0}{r}\right)^4\right]^{1/2},
\end{equation}
and $\tilde r_0:=\lambda^{-1}\sqrt{\frac{|\mu|}{4\pi}}$. Note that despite the fact that the monopole magnetic field and the point-charge electric field have different functional form, they both give rise to the same kind of effective geometry and hence the dynamics of null geodesics are entirely analogous. This equivalence of the geometries for the purely electric and magnetic backgrounds might have been anticipated on the basis of the electric-magnetic duality invariance of Born-Infeld theory. The fact that purely electric and purely magnetic backgrounds give rise to the same effective geometry encourages us to consider backgrounds with both electric and magnetic charge. This is the task we will undertake in the next section.

\section{Dyon}\label{sec dyon}
The existence of dyon-like solutions carrying both electric and magnetic charges in Born-Infeld electromagnetism has been shown in e.g. \cite{PhysRevD.61.085014}. In order to obtain the background solutions, we simply need to consider the magnetic monopole given by \eqref{monopole}, that trivially solves the field equations, and plug it into the field equations that will then determine the electric field background. The solution is given by
\begin{equation} \label{campoelecdyon}
    \Vec{E}=\frac{q}{4\pi r_{\text{D}}^2}\frac{1}{\sqrt{1+(r/r_{\text{D}})^4}}\hat{r},
\end{equation}
where the screening radius is now given by\footnote{In this section, we will set $\lambda=1$ to alleviate the notation so all the quantities will be measured in units of $\lambda$.}
\be
r_{\text{D}}^4=r_0^4+\tilde{r}_0^4=\frac{q^2+\mu^2}{(4\pi)^2}.
\ee
The dyon solution clearly has $Z\neq0$, so we need to employ the expressions for the effective metric for arbitrary non-vanishing $Z$. At this point, it should not come as a surprise that Born-Infeld electromagnetism is sufficiently amicable as to satisfy $\Delta=0$ also for $Z\neq0$ so the dyon background also fulfils the requirement to have one single effective metric. A difference with respect to the situation with $Z=0$ is that Eq. \eqref{KYZ} does not seem to be trivially satisfied. However, Born-Infeld theory again identically satisfies that equation. To see that this is a non-trivial statement, let us notice that we can obtain $\mathcal{B}_1\mathcal{B}_2$ from \eqref{KYZ} and \eqref{B2} so we have the relation
\be
\mathcal{B}_1\mathcal{B}_2=\frac{K_{YZ}}{Z}=\frac{K_{YY}-K_{ZZ}}{2Y}.
\ee
This is a non-trivial constraint on the derivatives of $K$ that is in turn satisfied by the Born-Infeld Lagrangian. Thus, the dyon background also admits an effective metric for the propagation of rays in the eikonal approximation given by
\be
g^{\mu\nu}=\eta^{\mu\nu}+\frac{1}{1-2Y}F^{\mu\rho}F_\rho{}^\nu,
\ee
where we have already substituted 
\be
\frac{K_{YY}-K_{ZZ}}{2Y}=\frac{1}{1-2Y}.
\ee
The line element (up to a conformal factor) can be written as
\begin{equation}
    \diff s^2= -\diff t^2+\diff r^2+ h_\text{D}^2(r)\diff \Omega^2,
\end{equation}
where the area function takes the form
\begin{equation}
     h^2_\text{D}(r)=r^2\frac{1+\vec{B}^2}{1-\vec{E}^2},
     \label{eq:hdyonEB}
\end{equation}
and, if we substitute the expressions for the magnetic and electric fields of the dyon, we obtain once again the expression
\begin{equation}
     h_D(r)=r\left[1+\left(\frac{r_\text{D}}{r}\right)^4\right]^{1/2},
\end{equation}
i.e., the wormhole geometry actually corresponds to a dyon with arbitrary electric and magnetic charge. This strongly suggests that the effective geometry is determined by the duality invariance of the Born-Infeld Lagrangian. To see this more clearly, we can express the area function in terms of the magnetic field $\vec{B}$ and the electric displacement $\vec{D}$ that is given by
\be
\vec{D}=\frac{\partial K}{\partial \vec{E}}=K_Y\vec{E}+K_Z\vec{B}=-\frac{\vec{E}+Z\vec{B}}{\sqrt{1-2Y-Z^2}}.
\label{eq:DtoE}
\ee
Let us notice that, since the electric displacement satisfies the usual equation for a static pointlike electric charge
\be
\nabla\cdot\vec{D}=q\delta(\Vec r),
\ee
its solution is simply
\begin{equation}
    \vec{D}=\frac{q}{4\pi r^2}\hat{r}.
\end{equation}
On the other hand, we can solve \eqref{eq:DtoE} for $\vec{E}$ in terms of $\vec{D}$ and $\vec{B}$, resulting
\be
\vec{E}=\frac{\vec{D}}{1+\vec{B}^2+\vec{D}^2}.
\ee
If we plug this relation into \eqref{eq:hdyonEB}, we obtain
\be
h^2_\text{D}(r)=r^2 \left(1+\vec{B}^2+\vec{D}^2\right).
\ee
Now, we recall that duality invariance can be realized as a rotation of the complex vector \mbox{$\vec{D}+i\vec{B}\to e^{i\alpha}\left(\vec{D}+i\vec{B}\right)$} for an arbitrary angle $\alpha$. We can finally corroborate that the area function is indeed duality invariant since it can be written as
\be
h^2_\text{D}(r)=r^2 \left(1+\left\vert \vec{D}+i\vec{B}\right\vert^2\right),
\ee
that is manifestly duality invariant. This provides a clear explanation for why the electric charge, the magnetic monopole and the dyon all give rise to the same effective wormhole geometry as a consequence of duality invariance.

\section{A digression on the scalar Dirac-Born-Infeld analogue}\label{Sec DBI}

The existence of remarkable properties that single out Born-Infeld electromagnetism  among the family of nonlinear electrodynamics also occurs for its scalar field counterpart, namely the Dirac-Born-Infeld (DBI) theory. If we consider the Lagrangian for a generic shift-symmetric scalar field
\be
\Lag=\mathcal{P}(X),\qquad X\equiv\frac12(\partial\varphi)^2,
\ee
the particular case of DBI that corresponds to the choice
\be
\mathcal{P}(X)=\Lambda^4\left(\sqrt{1+2X/\Lambda^4}-1\right),
\ee
exhibits unique properties that make it very special among the generic shift-symmetric theories (see e.g. \cite{Deser:1998wv,Mukohyama:2016ipl,deRham:2016ged,Pajer:2018egx,Grall:2019qof,BeltranJimenez:2024zmd}). Among such properties, we can mention that it has causal propagation and it avoids the formation of caustics \cite{Deser:1998wv,Mukohyama:2016ipl,deRham:2016ged}, very much like the Born-Infeld theory. The resemblance of the DBI and Born-Infeld Lagrangians is apparent so it does not come as a surprise that they share some remarkable properties.

Given the similarities between both theories, it is pertinent to wonder if the DBI theory shares the effective wormhole geometry for perturbations around a non-trivial background generated by a point-like particle similarly to the Born-Infeld theory. Furthermore, the question is relevant because the linear perturbations of the DBI field (and, in fact, for any $\mathcal{P}(X)$) can be exactly described in terms of an effective geometry \cite{Babichev:2007dw} and not only in the eikonal approximation as it occurs for Born-Infeld electromagnetism. The answer to the question is negative and this can be understood by simply considering the propagation speed of the perturbations. The DBI profile generated by a point-like particle has a gradient that exactly coincides with the Born-Infeld solution \eqref{campoelec} so we have
\begin{equation} 
    \nabla\varphi=\frac{q_s}{4\pi r_0^2}\frac{1}{\sqrt{1+(r/r_0)^4}}\hat{r},
\end{equation}
with $q_s$ the scalar charge of the particle and $r_0$ the screening radius analogous to the Born-Infeld case. The propagation speed for the scalar field perturbations around this background is given by
\be
c_s^2=1+\frac{2X\mathcal{P}_{XX}}{\mathcal{P}_X}=1+\frac{r_0^4}{r^4}.
\ee
At large distances, we recover $c_s^2\simeq 1$, but as the perturbation enters the screened region, the propagation speed becomes superluminal and eventually diverges at the position of the particle. This is a radically different situation from the Born-Infeld theory, where the perturbations propagate subluminally in the screened region and arbitrarily slowly as they approach the position of the particle. This slow down of the Born-Infeld perturbations permitted us to interpret why the effective geometry (in the eikonal approximation) could be associated with a wormhole. For the DBI theory, having the opposite behaviour, i.e., the perturbations propagate arbitrarily fast as they approach the position of the particle, the perturbations will not see an effective wormhole geometry. 

This conclusion can be made more explicit by considering the effective metric seen by the scalar perturbations that is given by \cite{Babichev:2007dw}
\be
g_s^{\mu\nu}=\mathcal{P}_X\eta^{\mu\nu}+ \mathcal{P}_{XX}\partial^\mu\varphi\partial^\nu\varphi.
\ee
For the considered background profile, the effective line element can be written as
\be
\mathcal{P}_X\diff s^2=-\diff t^2+\frac{\diff r^2}{c_s^{2}(r)}+r^2\diff\Omega^2.
\ee
If we transform to proper distance by the change of radial coordinate $\diff \rho=\diff r/c_s(r)$ we can put the metric in terms of an area function (up to the conformal factor $\mathcal{P}_X$) like in the Born-Infeld case. The area function $A=4\pi r^2(\rho)$ however does not present a minimum in the DBI case and, in fact, it goes to zero as $\rho^{2/3}$ at the position of the particle instead of diverging. Thus, the geometry does not correspond to a wormhole as in the Born-Infeld case. We will not perform a detailed analysis of the resulting effective geometry (which is beyond the scope of this work), but we will simply report the radically different nature of the effective geometries for these closely related theories.

\section{Conclusions}\label{conclusions}

In this work, we have studied nonlinear electrodynamics through which it is possible to formulate analogue models of gravity. To this end, we have computed the quadratic Lagrangian for the perturbations around a non-trivial background and obtained necessary conditions for the dynamics to be described by an effective metric. An important feature of nonlinear electromagnetism is that it is not possible, in general, to describe the entire dynamics of the perturbations in terms of a metric, but only in the eikonal approximation. This is an important difference with respect to nonlinear scalar field theories where the dynamics of the linear perturbations around non-trivial backgrounds admits an effective metric description for all frequencies (at linear order in perturbations). We have then obtained the effective metric for the propagation of rays in nonlinear electromagnetism for the cases when this is possible, up to a conformal factor that is irrelevant because rays follow null geodesics of the effective metric. This result is the basis for the construction of analogue models with nonlinear electromagnetism.

We have shown that Born-Infeld electromagnetism admits an effective metric, and we have shown that such an effective metric for a charged point-like particle describes a wormhole geometry. We have carried out a complete analysis of the light trajectories as a function of the impact parameter in the effective geometry. By numerically solving the equation of null geodesics we observed that, depending on the impact parameter of the geodesics, light rays experience a lensing effect, fall through the throat of the wormhole, or undergo an unstable circular trajectory in the throat of the wormhole. We have also seen how the behaviour of geodesics relates to the propagation velocity in the wormhole. This same dynamics has been obtained in a magnetic background created by a magnetic monopole and in the background generated by a dyon. In all cases, we obtain the same wormhole geometry and we have associated it with the duality invariance of the Born-Infeld theory.

Finally, we have briefly considered the scalar field version of Born-Infeld theory, i.e., the shift-symmetric DBI theory, to compare with the effective geometry generated by a particle with scalar charge. We have seen that, despite the great resemblance between the scalar and electromagnetic theories that makes them share exceptional properties, the effective geometries differ. In particular, the effective geometry in the DBI theory does not correspond to a wormhole. We have related the difference to the different behaviour of the propagation speed of the perturbations as they enter the strong field (screened) region. While in Born-Infeld theory, the propagation speed of the perturbations becomes smaller as they approach the origin, in DBI the opposite occurs, i.e., the propagation speed becomes increasingly larger as the trajectories approach the particle.

\vspace{6pt} 


\authorcontributions{Conceptualization: J.B.J., L.J.G., and M.P.G.; writing, original draft preparation: M.P.G.; writing, review and editing: J.B.J., L.J.G., and M.P.G.; all authors have read and agreed to the published version of the manuscript.}

\funding{
Financial support was provided by the Spanish Government through the Grants No. PID2020-118159GB-C44, PID2021-122938NB-I00, and PID2023-149018NB-C44 (funded by \linebreak MCIN/AEI/10.13039/501100011033). L.J.G. also acknowledges the support of the Natural Sciences and Engineering Research Council of Canada (NSERC). J.B.J. also acknowledges the support of the Project SA097P24 funded by Junta de Castilla y Le\'on.
}
   
\dataavailability{No new data were created or analysed in this study.}

\conflictsofinterest{The authors declare no conflicts of interest.} 


\begin{adjustwidth}{-\extralength}{0cm} 

\reftitle{References}

\bibliography{universe.bbl}
 
\PublishersNote{}
\end{adjustwidth}
\end{document}